\documentstyle{l-aa}

\newcommand{\lsim}{\raisebox{0.3mm}{\em $\, <$} \hspace{-3.3mm} 
\raisebox{-1.8mm}{\em $\sim \,$}}

\begin{document}

%

\thesaurus{03         
           (11.05.1;  
            11.05.2;  
            11.06.1;  
            11.16.1;  
            11.19.5)} 

\title{Origin of the colour-magnitude relation of elliptical galaxies}

\author{Tadayuki Kodama$^1$ \& Nobuo Arimoto$^{1,2}$}

\offprints{T. Kodama}

\institute{$^1$ Institute of Astronomy, University of Tokyo, Mitaka,
           Tokyo 181, Japan \\
           $^2$ Institute of Astronomy, University of Cambridge,
           Madingley Road, Cambridge, CB3 0HA, UK \\
           e-mail: kodama@mtk.ioa.s.u-tokyo.ac.jp}

\date{Received -----  --, 1996; accepted September 13, 1996}

\maketitle

\begin{abstract}

 Evolutionary models of elliptical galaxies are constructed by using
 a new population synthesis code. Model parameters are calibrated to
 reproduce the observed colour-magnitude (CM) relation of Coma ellipticals
 in $V-K$ vs. $M_{V}$ diagram.
 The SEDs are degenerated in stellar age and metallicity. An attempt
 is performed to break this degeneracy, by simulating 
 evolution of the CM relation of elliptical galaxies, 
 based on the two alternative interpretations; {\it i.e.,} the CM relation
 reflects different mean stellar age or various stellar metallicity.
 A confrontation with the CM diagrams of E/S0
 galaxies in the two distant clusters Abell 2390 ($z=0.228$) and
 Abell 851 ($z=0.407$) reinsures previous contentions that the CM
 relation is primarily a metallicity effect.
 This conclusion does not depend either on the model parameters, or on
 the cosmological parameters adopted.

\end{abstract}



\section{Introduction}

The integrated colours of elliptical galaxies become progressively bluer
towards fainter magnitudes (Faber 1973, 1977; Visvanathan \& Sandage 1977;
Frogel et al. 1978; Persson, Frogel, \& Aaronson 1979; Bower, Lucey, \&
Ellis 1992a; 1992b).
With an accurate photometry of a large sample of elliptical galaxies in Virgo
and Coma clusters of galaxies, Bower et al. (1992a, b) have
shown that the colour-magnitude (CM) relations of two clusters
 are identical within observational uncertainties, suggesting 
that the CM relation is universal 
for cluster ellipticals. The rms scatter
about the mean CM relation is typically $\sim 0.04$ mag, a
 comparable size to observational errors, and implies a virtually 
 negligible intrinsic scatter. If the scatter is due to age
 dispersion among cluster ellipticals, the photometric data by Bower et al.
 (1992a, b) suggest that cluster ellipticals are unlikely to have formed below
 a redshift of 2 ($q_{0}=0.5$), which sets an upper limit of $\sim 2$
 Gyrs for the age dispersion of the bulk stellar populations of cluster
 ellipticals. An identical upper limit to the age dispersion of such
 galaxies is also suggested by their small dispersion about the
 so-called fundamental plane (Renzini \& Ciotti 1993).
 These findings are in excellent agreement with a scenario for the
 formation of elliptical galaxies dominated by protogalactic collapse
 and dissipational star formation early in the evolution of the
 universe (Larson 1974; 
 Saito 1979a, b; Arimoto \& Yoshii 1986, 1987, hereafter AY87; 
 Matteucci \& Tornamb\'e 1987; Yoshii \& Arimoto 1987; Bressan, Chiosi, \&
 Fagotto, 1994; Tantalo et al. 1996).
 The CM relation itself is also quite naturally established in
 collapse/wind model of galaxy formation; a galactic wind is
 induced progressively later in more massive galaxies owing to
 deeper potential and stellar populations in brighter galaxies
 are much more enhanced in heavy elements and as a result redder in colours.

 This {\it conventional} interpretation of the CM relation of ellipticals,
 however, has been questioned by Worthey (1996) who points out, based on
 his population synthesis model (Worthey 1994), that the sequence of
 colours and line strengths among ellipticals can be almost equally
 well explained by a progressive decrease of either a mean stellar 
 metallicity or an effective stellar age. 
 Recently, Bressan, Chiosi, \& Tantalo (1996) have performed a detailed
 analysis of the CM relation of elliptical galaxies. In addition to broad
 band colours, the authors have considered absorption line strengths and
 their radial gradients, abundance ratios, spectral energy distribution in
 UV-optical region, and stellar velocity dispersion, all of which might be
 tightly coupled with the CM relation itself, and have suggested that
 the history of star formation in elliptical galaxies has probably been
 more complicated and heterogeneous than the galactic wind scenario
 mentioned above.

 The dual interpretations of the CM relation are now well known as the
 {\it age-metallicity degeneracy} of old stellar populations.
 Arimoto (1996) have distributed the spectral templates of nearby
 ellipticals (E1 for giants and E4 for dwarfs; Bica 1988) to about
 dozen groups of population synthesis (E.Bica \& D.Alloin;
 A.Pickles; R.W.O'Connell \& B.Dorman; W.Kollatschny \& W.Goerdt;
 G.Bruzual \& S.Charlot; M.Fioc \& B.Rocca-Volmerange; G.Worthey;
 B.Poggianti \& G.Barbaro; U.Fritz-v.Alvensleben; J.-H.Park \& Y.-W.Lee;
 R.F.Peletier, A.Vazdekis, E.Casuso, \& J.E.Beckman; N.Arimoto \& T.Yamada),
 both  evolutionary and optimizing, and 
 have asked to report their best fit solutions.
 The results are surprising: for E1 exercise most groups come to the
 same conclusion; {\it i.e.,} nuclei of giant ellipticals are old (10-19) Gyrs
 and mean metallicities are at least solar or more.
 However, for E4
 exercise, they do not agree at all. They find that nuclei of
 dwarf ellipticals are either 1) old ($11-15$) Gyrs and metal-rich
 ($2.5-3Z_{\odot}$), or 2) old ($10-16$) Gyrs and solar metallicity
 or metal-poor ($0.15-0.7Z_{\odot}$), or 3) young ($3-5$) Gyrs
 and solar metallicity or metal-rich ($2Z_{\odot}$)
[see Arimoto (1996) for more details]. 
 This confusing results for E4 exercise come 
 from the age-metallicity degeneracy and not from different approach
 (evolutionary or optimizing),
 nor from different techniques and ingredients adopted in the models.
 
 All in all, the previous interpretation that the CM relation is primarily
 a metallicity effect looks less secured today. It is not clear anymore 
 whether fainter ellipticals are bluer because stellar populations are 
 younger or because stars are more metal-deficient in average.
 Indeed an alternative scenario for the formation of elliptical
 galaxies suggests that ellipticals result from the merging of
 lesser stellar systems taking place mostly at later times and
 involving substantial star formation (Toomre \& Toomre 1972;
 Schweizer \& Seitzer 1992; Fritz-von Alvensleben \& Gerhard 1994). 
 This scenario is in good agreement with fine structures and kinematic
 anomalies of some elliptical galaxies (eg., Kormendy \& Djorgovski 1989).
 In addition to this, the optimizing synthesis tend to pick up finite
 fraction of light from intermediate age populations, which also supports
 recent episodes of star formation in ellipticals (O'Connell 1976;
 Pickles 1985b; Rose 1985; O'Connell 1986; Bica 1988; see also Arimoto 1996).

 Although several attempts have been made to disentangle the age and 
 metallicity effects on the line strengths as well as on the broad band 
 colours (Buzzoni, Gariboldi, \& Mantegazza 1992; Buzzoni, Chincarini, 
 \& Molinari 1993; Gonzales 1993; Worthey 1994; Worthey et al. 1994),
 it is still very difficult to break the age-metallicity degeneracy of
 elliptical galaxies so far as one sticks to photo-spectroscopic data of
 galaxies at the present epoch. However, it is comparatively easier to
 show that the CM relation is indeed primarily due to the stellar metallicity
 effect if one studies evolution of the CM relation as a function of 
 look-back time. If the CM relation is an age sequence, it should evolve
 rapidly and should disappear beyond certain redshift, because fainter
 galaxies are approaching to their formation epoch. Contrary, if the CM 
 relation is a metallicity sequence and 
 ellipticals are essentially old, it should evolve
 {\it passively} and should be 
 still traced at significantly high redshifts.
 In this paper, we compare the theoretical evolution of the CM relation
 with the CM relations of cluster ellipticals at cosmological distances
 ($z \simeq 0.2-0.4$) and show 
 that the bulk of stars were probably formed early in elliptical
 galaxies,
 and that the CM relation takes its origin at early times from galactic wind
 feedback, thus reinsuring confirmation of previous contentions.

 In \S2, we outline our population synthesis model,
 and in \S3 we present the theoretical evolution of the CM relation
 and confront it with the CM diagrams of elliptical/S0 galaxies of two
 distant clusters Abell 2390 ($z=0.228$) and Abell 851 ($z=0.407$).
 We discuss implications of several effects, such as 
 dynamical disturbance, intermediate age stars, and new star formation,
 on the formation of elliptical galaxies in \S4.
 Our conclusions are given in \S5.

\section{Model}

\subsection{Stellar population synthesis code}

 The basic structure of our new code follows Arimoto \& Yoshii's
 (1986) stellar population synthesis prescription that takes into account the
 effects of stellar metallicity on integrated colours of galaxies
 for the first time.  New stellar evolutionary tracks are incorporated
 comprehensively, and late stellar
 evolutionary stages are fully taken into account.
 The code uses the so-called {\it isochrone synthesis} technique
 (cf. Charlot \& Bruzual 1991) and gives the evolution of synthesized
 spectra of galaxies in a consistent manner with galaxy chemical evolution,
 and the details of the code
 will appear in Kodama \& Arimoto (1996). A brief description for
 essential techniques and the ingredients are given below (see also
 Table 1).

\subsubsection{Stellar evolutionary tracks}

 {\it Original tracks}: By using the stellar evolution code provided
 by Saio \& Nomoto (1994; private communication),
 we have calculated a grid of stellar evolutionary
 tracks for low mass stars $(0.6 \le m/M_{\odot} \lsim 2.4)$
 from zero-age main sequence (ZAMS) to tip of red giant branch
 (RGB), and for intermediate mass and massive stars $(2.4 \lsim m/M_{\odot}
 \le 60)$ from ZAMS to onset of carbon ignition.
 These tracks are calculated for low stellar metallicities
 $Z=0.0001$, 0.0002, 0.0005, 0.001, and 0.002 with helium abundance
 $Y = 0.23 + 2.5 Z$. The revised radiative opacities by Iglesias (1993)
 are used and a mixing length parameter $\ell/H_p$ is set to
 be 1.5 throughout. For the RGB evolution, the mass loss law of Reimers (1977)
 with an efficiency parameter $\eta=1/3$ is assumed and those given by
 Nieuwenhuijzen \& de Jager (1990) are adopted for the rest of evolutionary
 stages. We have compared our stellar evolutionary tracks for $Z=0.0005$ 
 with those of Padua group (see below) for $Z=0.0004$. The appearance on the
 HR diagram is nearly identical for all masses, but the evolutionary 
 time of our tracks is systematically $\sim 10$\% longer than the
 Padua tracks. This is essentially due to a difference in the program structure
 and the input physics between the two codes, and we 
 do not try to adjust the evolutionary time of our tracks.

 {\it Padua tracks}: For stars of higher metallicities
 $Z=0.004$, 0.008, 0.02, and 0.05,
 we adopt Padua stellar evolutionary tracks from the ZAMS to the tip of
 RGB for low mass stars 
 $(0.6 \le m/M_{\odot} \lsim 2.4)$ and to the onset of carbon ignition
 to the rest
 $(2.4 \lsim m/M_{\odot} \le 60)$, with $Y = 0.23 + 2.5 Z$
 (Bressan et al. 1993; Fagotto et al. 1994a; 1994b). Padua tracks are
 calculated with the revised radiative opacities by Iglesias et al.
 (1992) and with $\ell/H_p = 1.63$. Stellar mass loss is considered for
 massive stars $(m/M_{\odot} \ge 12)$.

 {\it Lower main sequence}: Lower main sequence stars
 $(0.1 \le m/M_{\odot} \le 0.5)$ are assumed to stay on the ZAMS over
 the Hubble time. Tracks are taken from VandenBerg (1983) for
 $0.0001 \le Z \le 0.02$ and are extrapolated for $Z=0.05$.
 $Y=0.25$ is adopted regardless of $Z$. Opacities are adopted
 from Cox \& Stewart (1970a; 1970b), Cox \& Tabor (1976) and Alexander (1975).
 A low value of $\ell/H_p = 1.0$ is assumed
 but the location of ZAMS on the HR diagram for lower main sequence
 stars is rather insensitive to a mixing length parameter since
 convective energy is transported almost adiabatically in these stars
 (VandenBerg 1983).

 {\it Horizontal branch}: We adopt the revised Yale tracks for
 low mass horizontal branch (HB) evolution which are kindly
 provided by S.Yi \& P.Demarque (1995; private communication).
 The tracks are calculated with the revised OPAL opacities
 (Iglesias \& Rogers 1991; Rogers \& Iglesias 1992)
 and $\ell/H_p = 1.7$ for a full range of $Z$ and $Y$.
 For our use, 9 sets of HB tracks are interpolated
 from the original grid for $Z=0.0001$ to $0.05$ and $Y$ at the ZAMS.
 For details of the input physics, see Yi (1996).
 HB stars are distributed on the HR diagram
 by using the same method developed by Rood (1973) with a
 total mass dispersion $\sigma_{M} = 0.025 M_{\odot}$.

 {\it Asymptotic giant branch, post-AGB, and white dwarf}:
 Evolutionary tracks of low mass asymptotic giant branch (AGB) stars are
 taken from Vassiliadis \& Wood (1993) and those of post-AGB stars
 (H-burning tracks) and white dwarfs are taken from Vassiliadis \&
 Wood (1994). These tracks are calculated for $0.001 \le Z \le 0.02$,
 thus tracks for
 lower and higher metallicities are extrapolated.
 $Y=0.25$ is assumed throughout.
 Opacities are taken from Huebner (1977) and Bessell et al. (1989).
 The mixing length parameter is set to be $\ell/H_p = 1.6$.
 A formula of  mass loss rate is applied for AGB stars according to the
 empirical relation between mass loss rate
 and pulsational period.

\subsubsection{Stellar spectra}

 Kurucz's (1992) stellar flux library is adopted for the metallicity
 $0.0001 \le Z \le 0.05$, the effective temperature
 $4000K \le T_{\rm eff} \le 50000K$, and the surface
 gravity $0.0 \le \log g \le 5.0$.  The library covers the wavelength range
 from UV to near IR with spectral resolutions $\Delta \lambda =
 10-20\AA$ for $\lambda \leq 1\mu m$ and
 $\Delta \lambda = 50-100\AA$ for $\lambda \ge 1\mu m$.
 We interpolate linearly a logarithm of the flux $F_{\lambda}$
 in $\log Z$, $\log T_{\rm eff}$ and $\log g$, if necessary.

 Below 4000$K$, empirical stellar spectral libraries of Pickles (1985a) are
 employed with a supplemental use of  Gunn \& Stryker (1983).
 The former covers the wavelength range from
 $\lambda = 3600\AA$ to $10000\AA$
 with the resolution $\Delta \lambda = 10-17\AA$, while
 the latter covers from $\lambda = 3130\AA$ to $10800\AA$
 with the resolution $\Delta \lambda = 20-40\AA$.
 Fluxes are extended towards the
 near IR region using empirical colours
 in $JHKL$ bands for M stars by Bessell \& Brett (1988):
 Effective temperatures are allocated
 according to Ridgway (1980) and Pickles (1985a).
 Since spectra of M stars with non-solar metallicity are hardly
 available, the line blanketing effects of metallicity are ignored.
 This could be justified partly by the fact that there are very few M
 giants with $Z \le 1/3 Z_{\odot}$. Bolometric corrections for
 M stars are taken from Bessell \& Wood (1984) and Johnson (1966).

 Above 50000$K$, spectra are almost featureless, and black body spectra are
 assumed.

\subsubsection{Simple stellar population (SSP) model}

 SSP is defined as a single generation of coeval and chemically
 homogeneous stars of various masses (Renzini \& Buzzoni 1986; Buzzoni 1989).
 The SSP is a building block of galactic population synthesis model and any
 stellar system of composite age and metallicity, like a galaxy,
 can be constructed as a linear combination of the SSPs
 (Charlot \& Bruzual 1991; Bruzual \& Charlot 1993).
 SSP spectrum is constructed as follows.
 Isochrone is first constructed from the stellar evolutionary tracks for
 the specified age and metallicity. Stellar spectra are assigned for
 each point on the isochrone and the SSP spectrum is synthesized by
 integrating stellar spectra along the isochrone with the number of
 stars, estimated from the initial mass function (IMF), as a weight.

 Broad band colours are calculated by applying the filter response
 functions to the SSP spectrum.
 The resulting colours are confronted with the observed colours
 of star clusters in the Milky Way (Harris 1996) and the Magellanic Clouds
 (van den Berg 1981) and are confirmed to reproduce the observed ones.
 The $B-V$ evolution of the solar metallicity SSPs is
 nearly identical to that of Bruzual \& Charlot (1995), while the
 $V-K$ are slightly redder than Bruzual \& Charlot (1995)
 for the SSPs older than 13 Gyrs and approach progressively towards 
 Bertelli et al.'s (1994; see Charlot, Worthey, \& Bressan 1996).

 \begin{table*}
   \begin{center}
   \caption{Characteristics of Stellar Population Synthesis Code.}
   \label{table-1}
   \begin{tabular}{ll}
   \hline
   Stellar evolutionary tracks & this work,\\
    & Bressan et al. (1994), Fagotto et al. (1994a,b),\\
    & VandenBerg (1983), Yi \& Demarque (1995),\\
    & Vassiliadis \& Wood (1993,1994)\\
   Interior opacities & Iglesias et al. (1992), Iglesias (1993)\\
   Mixing length ($\ell/H_p$) & $1.5-1.7$\\
   Mass loss rate & Nieuwenhuijzen \& de Jager (1990),\\
    & Reimers (1977) ($\eta=1/3$),\\
    & empirical formula (AGB)\\
   Flux library & Kurucz (1992)\\
   M giants spectra  & Pickles (1985a), Gunn \& Stryker (1983)\\
   \hspace*{1.22cm} near IR colours & Bessell \& Brett (1988)\\
   \hspace*{1.22cm} effective temperature & Ridgway et al. (1980)\\
   \hspace*{1.22cm} bolometric correction & Bessell \& Wood (1984)\\
   Metallicity range & $0.0001 \le Z \le 0.05$\\
   Mass range & $0.1 \le m/M_{\odot} \le 60$\\
   \hline
   \end{tabular}
   \end{center}
 \end{table*}

\subsubsection{Chemical evolution}

 The stellar populations of any galactic model can be described by
 a linear combination of the SSP models, provided that a mass weight
 of each SSP model specified with a single set of age and metallicity
 is known {\it a priori}. The fractional contributions of SSP models
 should come from the detailed history of star formation and
 nucleosynthesis, or in other words, a modelling of galactic chemical
 evolution is of vital importance in synthesizing a galaxy spectrum.

 Since our major concern in the present study
 is the colour evolution of elliptical galaxies, we hereafter
 consider the so-called infall model instead of the simple model
 of galaxy chemical evolution (see \S 2.2).
 We assume that the galactic gas is supplied from the
 surrounding gas reservoir trapped in the gravitational
 potential of a galaxy, and that the gas is always well-mixed and
 distributes uniformly.
 Then time variations of the gas mass $M_{g}$ and the gas metallicity
 $Z_{g}$ are given by the following equations (Tinsley 1980):
 \begin{equation}
 \frac{dM_{g}(t)}{dt} = - \psi(t) + E(t) + \xi_{in}(t),
 \end{equation}
 \begin{equation}
 \frac{dZ_{g}(t)}{dt} = \frac{1}{M_{g}(t)} \left\{ E_{Z}(t) - \psi(t)Z_{g}(t) +
\xi_{in}(t)Z_{in}(t) \right\},
 \end{equation}
 where $\psi(t)$ and $\xi_{in}(t)$ are
 the star formation rate (SFR) and the gas infall rate, respectively.
 As for the initial conditions, we assume that there was no gas in a
 galaxy before the onset of star formation; ie., $M_{g}(0)=0$ and $Z_{g}(0)=0$.
 The SFR $\psi(t)$ is assumed to be proportional to the gas mass
 (Schmidt law):
 \begin{equation}
 \psi(t) = \frac{1}{\tau} M_{g}(t),
 \end{equation}
 where $\tau$ gives the star formation time scale in Gyrs.
 $\xi_{in}(t)$ is assumed to be expressed with the initial total mass of
 gas reservoir $M_{T}$ and the gas infall time scale $\tau_{in}$ as,
 \begin{equation}
 \xi_{in}(t) =  \frac{M_T}{\tau_{in}} \exp(-\frac{t}{\tau_{in}})
 \end{equation}
 (cf. K\"oppen \& Arimoto 1990).
 $Z_{in}$ is the metallicity of the infalling gas and assumed to be zero.
 $E(t)$ and $E_{Z}(t)$ in Eqs.(1)-(2)
 are the ejection rate of the gas and
 the metals from dying
 stars, respectively, and are calculated from the following integrals:
 \begin{equation}
 E(t) = \int_{m_{t}}^{m_{u}} (1-w_{m}) \psi(t-\tau_{m}) \phi(m) dm,
 \end{equation}
 \begin{eqnarray}
 E_{z}(t) = \int_{m_{t}}^{m_{u}} & \{(1-w_{m}-p_{zm}) Z_{g}(t-\tau_{m}) +p_{zm}\} \nonumber \\
 & \times \psi(t-\tau_{m}) \phi(m) dm,
 \end{eqnarray}
 where $w_{m}$ and $\tau_{m}$ are the remnant mass fraction and the lifetime
 of stars with mass $m$, respectively.
 $p_{zm}$ is the mass fraction of newly synthesized
 and ejected metals.  Lower mass limit $m_{t}$ for both integrals
 is the stellar mass with lifetime $\tau_{m}=t$.
 Nucleosynthesis data $w_{m}$ and $p_{zm}$ are taken from Nomoto (1993;
 private communication)
 for $m \geq 10 M_{\odot}$ and from Renzini \& Voli (1981) for
 $m < 10 M_{\odot}$.

 The IMF $\phi(m)$ is defined by mass fraction and assumed to have
 a single power law of mass:
 \begin{equation}
 \phi(m)dm = A m^{-x} dm,\hspace{0.5cm} m_{l} \le m \le m_{u},
 \end{equation}
 where $m_{l}$ and $m_{u}$ are lower and upper limits of
 initial stellar mass, respectively. The Salpeter (1955) mass function
 has a slope $x = 1.35$ in this definition.
 The coefficient $A$ is determined by,
 \begin{equation}
 \int_{m_{l}}^{m_{u}} \phi(m)dm = 1.
 \end{equation}
 The IMF is assumed to be time invariant.

 Recent chemical evolution analyses of disc dwarfs and halo giants
 in the solar neighbourhood suggest that the mean lifetime of
 progenitors of type Ia supernovae (SNIa) that are responsible for
 iron production is not an order of $10^8$ yrs as previously believed
 but as long as $(1.5-2.5)$ Gyrs (Ishimaru \& Arimoto 1996;
 Yoshii, Tsujimoto, \& Nomoto 1996). This implies that SNIa's
 start to explode longer after when the bulk of stars formed in
 ellipticals as we will see in \S 2.2. Therefore, we ignore SNIa
 in this study.

 Finally, the synthesized spectrum $F_{\lambda}$ of a model
 as a function of galaxy age $T$ can be given by,
 \begin{equation}
  F_{\lambda}(T) = \int_{0}^{T} \psi(t)f_{\lambda}(T-t,Z(t)) dt,
 \end{equation}
 where $f_{\lambda}(T-t,Z(t))$ is the SSP spectrum with age $T-t$ and
 metallicity $Z(t)$.

\subsection{Evolution model of elliptical galaxies}

 A formation picture of elliptical galaxies by Larson (1974)
 and AY87 assumes that initial bursts of star formation occurred
 during rapidly infalling stage of proto-galactic gas clouds
 and that the gas was expelled when the cumulative thermal energy
 from supernova explosions exceeded the binding energy of the
 remaining gas. No stars formed afterwards and a galaxy has
 evolved passively since then.

 Following AY87, we assume that star formation abruptly terminates
 at the epoch of galactic wind $t=t_{gw}$; {\it i.e.},
 \begin{equation}
 \psi(t) = 0 \hspace{0.2cm} for \hspace{0.2cm} t \geq t_{gw}.
 \end{equation}
 A precise value of $t_{gw}$ is rather difficult
 to evaluate, because it would depend strongly on how much dark matter is
 involved in defining a gravitational potential as well as on the
 change of galactic radius during the gravitationally collapsing
 phase. Instead, we determine $t_{gw}$ empirically
 in such a way that models can reproduce $V-K$ 
 of nearby elliptical galaxies in Coma cluster.
 Open circles in Figs.1a and 1b give the CM relations of Coma ellipticals
 in $V-K$ vs. $M_{V}$ and $U-V$ vs. $M_{V}$ diagram, respectively
 (Bower et al. 1992a). Observational errors are 
 estimated to be 0.035 mag and 0.037 mag
 in $V-K$ and $U-V$, respectively. Absolute magnitudes $M_{V}$ are
 calculated with a help of the distance modulus of Coma cluster
 $(m-M)_{0}=34.6$.
 This value is obtained from the relative distance modulus between Coma and
 Virgo, $\Delta(m-M)_{0}=3.6$ (Bower et al. 1992b), and Tully-Fisher distance
 of Virgo core $(m-M)_{0}=31.0$ (Jacoby et al. 1992).
 Two filled squares indicate Bica's (1988) E1 and E4 spectral templates
 of nearby ellipticals adapted from Arimoto (1996).

 \begin{figure}
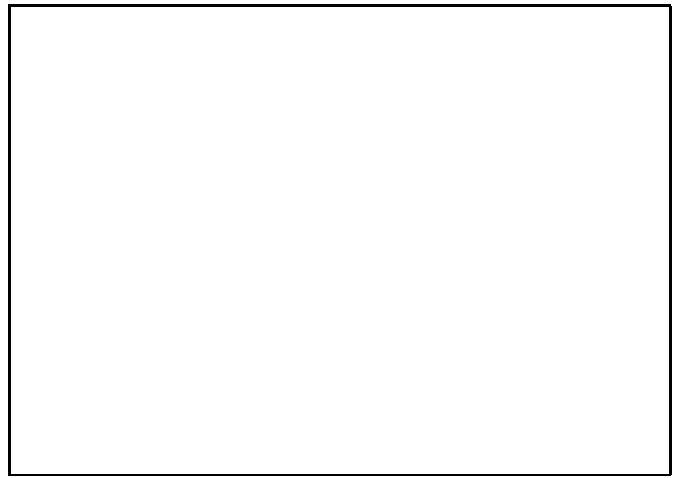

 \picplace{6.23cm}
 \caption {(a) CM relation of Coma ellipticals in $V-K$ vs. $M_V$
 diagram. Open circles are Coma ellipticals and a regression line
 is given by a solid line taken from Bower et al.(1992a,b) (BLE92).
 Filled squares gives Bica's (1988) E1 and E4 templates.
 Dashed and dotted lines represent loci of models defined as
 the metallicity sequence and the age sequence, respectively.
 (b) The same as Fig.1a, but in $U-V$ vs. $M_V$ diagram.}
 \label{fig-1}
 \end{figure}

 In Figs.1 and 2, the CM relations can be clearly
 identified with surprisingly small scatter. It is now well recognized
 that the CM relations can be interpretated alternatively
 either by a decrease of mean stellar metallicity towards fainter galaxies
 (AY87; Matteucci \& Tornamb\`e 1987; Bressan, Chiosi \& Fagotto 1994)
 or by a decrease of mean stellar age towards dwarf ellipticals (Worthey 1996).
 This happens because metallicity difference $\Delta \log Z$ or age difference
 $\Delta \log$ Age in two old stellar populations gives almost identical
 change in any colours and spectral indices
 if $\Delta \log {\rm Age} / \Delta
 \log Z = -3/2$ is kept (The 3/2 rule of Worthey 1994).
 This is the so-called {\it age-metallicity degeneracy}
 for old stellar populations (cf. Arimoto 1996).

 To break the age-metallicity degeneracy, Arimoto (1996) suggests
 to verify if the theoretical CM relation, defined either by the metallicity
 or by the mean age, is consistent with the empirical CM relations of cluster
 ellipticals at cosmological distances. 
 Therefore, we have simulated the evolution of the CM relation of
 elliptical galaxies for the two alternative scenarios.
 The resulting CM relations are then confronted with observational data,
 in the observer's frame, of elliptical galaxies in two distant clusters
 Abell 2390 ($z=0.228$) and Abell 851 ($z=0.407$) (\S 3.2).

 We regard the CM relation
 in the $V-K$ vs. $M_{V}$ diagram of Coma ellipticals (Fig.1a)
 as a standard sequence of nearby ellipticals to which
 models for present day ellipticals should be fitted.
 Model parameters are chosen as $x=1.20$, $m_{l} = 0.1 M_{\odot}$,
 $m_{u} = 60 M_{\odot}$, $\tau = 0.1$ Gyrs, and $\tau_{in} = 0.1$ Gyrs.
 $m_{l} = 0.1 M_{\odot}$ is chosen to give a mass-to-light ratio
 $M/L_B \simeq 8$ for giant ellipticals (Faber \& Jackson 1976; Michard 1980;
 Schechter 1980). Although the choices of $x$, $\tau$, and $\tau_{in}$
 are somewhat arbitrary, we note that our analysis is rather insensitive
 to these parameters due to the following reasons: 1) Colour evolution
 of the old SSPs depends negligibly on $x$ (Tinsley 1980), because
 stars that determine the synthesized colours of the old SSPs are
 main sequence stars near the turnoff and red giants; mass difference
 among them is negligibly small. 2) Relative dispersion of stellar
 ages within a galaxy is very small in our formation picture of elliptical
 galaxies. 3) Since $t_{gw}$ is determined so as to
 reproduce $V-K$ at $z=0$ (in the case of metallicity
 sequence introduced below), the mean stellar metallicity of a galaxy is
 adjusted automatically for any set of $x$, $\tau$, $\tau_{in}$, and $T_G$.
 Thus, nearly identical stellar contents for each galaxy is realized
 in terms of the mean stellar age and the metallicity, even if different
 set of parameters is chosen.\\

 \noindent (a) {\it metallicity sequence} :
 The colour change along the CM relation in $V-K$ is explained simply by a
 difference of mean stellar metallicities of galaxies (AY87).
 Following this interpretation, we construct a sequence of models
 with different mean metallicities.
 All ellipticals are assumed to have an equivalent age $T_{G} = 15$ Gyrs.
 In galactic wind model, the metallicity difference
 can be produced by assigning different $t_{gw}$.
 A standard CM relation in $V-K$ is, then, reproduced with
 $t_{gw}=0.52$ Gyrs at $M_V=-23$ mag and $t_{gw}=0.10$ Gyrs at $M_V=-17$ mag.
 In total 13 models are calculated for 0.5 magnitude interval in $M_{V}$
 and properties of some models at the present epoch are summarized in
 the upper part of
 Table 2:  $M/L_B$ is given in solar units, galaxy mass $M_{G}$ is
 calculated from $M_{V}$ and $M/L_V$. Luminosity-weighted
 mean stellar metallicity is defined as:
 \begin{equation}
 <\log Z/Z_{\odot}> = \frac{\Sigma (\log Z/Z_{\odot})
 L_{\rm BOL}}{\Sigma L_{\rm BOL}},
 \end{equation}
 where summation is taken for all stars in a galaxy (cf. AY87).
 The solar metallicity $Z_{\odot}=0.019$ is taken from
 Anders \& Grevesse (1989).
 We note that $t_{gw}$ is always less than the lifetime of SNIa (cf. \S 2.1.4)
 and that $<\log Z/Z_{\odot}>$ ranges from $-$0.014 to $-$0.596 from the
 brightest end to the faintest end. Alternatively, the luminosity-weighted
 mean stellar metallicity can be defined as (Greggio 1996):
 \begin{equation}
 \log <Z/Z_{\odot}> =  \log  \left\{ \frac{\Sigma (Z/Z_{\odot})
 L_{\rm BOL}}{\Sigma L_{\rm BOL}} \right\}.
 \end{equation}
 The resulting metallicities are also given in Table 2 for comparison.
 Eq.(12) gives $\sim +0.15$ dex higher values than Eq.(11).\\

 \noindent (b) {\it age sequence} :
 The CM relation can be explained alternatively by a progressive decrease of
 galaxy age towards fainter ellipticals (Worthey 1996).
 This can be represented by a sequence of models with fixed $t_{gw}$
 and various galaxy ages. A wind time is fixed to be the same as
 the brightest model of the metallicity sequence; ie., $t_{gw} = 0.52$ Gyrs.
 A standard CM relation in $V-K$ is, then, reproduced with
 $T_G=15$ Gyrs at $M_V=-23$ mag and $T_G=2.3$ Gyrs at $M_V=-17$ mag.
 In total 25 models are calculated for 0.25 magnitude interval
 in $M_{V}$ and properties of some models at the present epoch are
 summarized in the lower part of Table 2.
 All models have $<\log Z/Z_{\odot}> \simeq 0$ in this sequence.\\

 Models given in Tables 2 are all calculated under the context of the
 infall model of galaxy chemical evolution. A simple model approach is
 also attempted, but a resulting 
 $U-V$ is bluer by 0.15 mag 
 than the observed CM relation of Coma
 ellipticals at brightest end ($M_{V}=-23$ mag), 
 while the model gives a good fit to $V-K$.
 This implies that the simple model
 predicts too many metal deficient stars (G-dwarf problem of elliptical
 galaxies). As is in the solar neighbourhood, 
 infall model is effective in solving the G-dwarf problem and
 the observable properties of infall model
 were extensively discussed by Arimoto \& Jablonka (1991).
 
 The metallicity sequence and the age sequence
 are shown in Figs.1a and 1b with dashed and dotted lines, respectively.
 The same filter systems as Bower et al. (1992a) are adopted.
 Since both sequences are calibrated to the CM relation in $V-K$ vs. $M_{V}$
 diagram, these two lines are identical to Bower et al.'s (1992b)
 regression (solid) line in Fig.1a.
 On the contrary, $U-V$ of the metallicity sequence are systematically
 redder than the regression line of Bower et al. (1992b), amounting 
 $\Delta (U-V) \simeq 0.06$ at the faintest end. Figure 15 of 
 Charlot et al. (1996) suggests that this is partly due to our use
 of Kurucz (1992) spectra, but it is also possible that this is due to the
 neglect of binary stars in our population synthesis code. Binary main
 sequence stars will give rise of blue strugglers which have brighter
 magnitudes than the turnoff stars, thus will increase the flux in U-band. 
 Alternatively, the $U-V$ discrepancy can be attributed to an aperture
 effect. Bower et al. (1992a; b) adopted the same small aperture size for all
 galaxies, and if elliptical galaxies have metallicity gradient
 decreasing outward (e.g., Davies, Sadler, \& Peletier, 1993),
 the fixed aperture photometry should get progressively more
 light from metal-poorer stars in smaller galaxies. However, such a 
 detailed modeling of 2D and/or 3D metallicity gradient is beyond a scope
 of the present study.   
 We therefore do not use $U-V$ to calibrate the models. 
 Conclusions of this work is entirely free from this $U-V$ mismatch.

Recently Bressan, Chiosi, and Tantalo (1996) have conducted a detailed 
study of stellar populations in elliptical galaxies. The authors also 
adopt the infall model of chemical evolution. 
Table 3 compares the models on the metallicity sequence of this work
with those of Bressan et al. (1996) at galaxy age $T_{G} = 15$ Gyrs.
The Salpeter IMF with 
$m_l\simeq 0.16M_{\odot}$ (we calculate this value from their parameter
$\zeta=0.5$ by assuming the turnoff mass $m_t=0.9M_{\odot}$)
and $m_u=120M_{\odot}$ and constant $\tau_{in} = 0.1$ are 
assumed in their models. 
The definitions and the units of model properties are the same as 
those given in Table 2. $\tau$ and $M/L_{B}$ of Bressan et al. (1996) are 
re-scaled to our definition. Bressan et al. define the mean stellar 
metallicity as:
\begin{equation}
 < Z > = \frac{\int_0^{T_G} \psi(t) Z(t) dt}{\int_0^{T_G} \psi(t) dt},
\end{equation}
which gives the mass-weighted mean metallicity of stars ever formed.
We therefore re-calculate $<Z>$ for the models of this study according to 
Eq.(12). The model comparison has been done for two typical galaxies
which have different mass along the CM relation.
Table 3 shows that chemical and photometric properties 
are almost similar between the two models.

Although it is not explicitly shown in Table 3, their model predicts
higher [Mg/Fe] ratios for more massive galaxies, while ours would give
virtually the same values. This is because their model assumes
shorter time scale of star formation for massive galaxies, while our model
assumes it constant for all galaxies. As a result, their model achieves
an earlier occurrence of galactic wind in more massive galaxies.
Indeed, there is an observational trend
that ellipticals with stronger Mg$_2$ indices tend to
have larger line strengths of iron (Fe5270 and Fe5335), 
but the increase in iron is less than the prediction given by a population 
synthesis model with solar [Mg/Fe] (Worthey et al. 1992). 
However, the scatter is large and real. Moreover, it is not clear if this
means an increase of [Mg/Fe] towards luminous ellipticals, because unlike
single star spectroscopy these lines come from many stars of different age, 
metallicity, and probably abundance ratios, and a population synthesis 
approach is required to interpret [Mg/Fe] from these line strengths. However,
the present day population synthesis models are not yet matured to conduct
such analysis because behaviours of neither stellar evolutionary tracks 
nor stellar model atmosphere are fully investigated with various elemental 
abundance ratios. Thus, we believe that the trend of [Mg/Fe] with galaxy
luminosity needs to be confirmed and that one need not necessary take it
as a strong constraint for modeling elliptical galaxies. We should also note
that the galactic wind time in Bressan et al.'s (1996) models is in
the range of $0.31-0.43$ Gyrs, this would give [Mg/Fe] in the
``right'' direction if the mean lifetime of SNIa progenitors is 0.25
Gyrs (Matteucci 1994). However, recent studies suggest that the mean
lifetime of SNIa progenitors that are most responsible to the iron 
enrichment could be as long as $(1.5-2.5)$ Gyrs
(Ishimaru \& Arimoto 1996; Yoshii et al. 1996). If that is the case,
much longer time scale of star formation would be required for less
massive galaxies than Bressan et al.'s (1996) models.

 \begin{table*}
   \begin{center}
   \caption{Model sequences of elliptical galaxies at $z = 0$.}
   \label{table-2}
   \begin{tabular}{c|l|rrrrrrr}\hline\hline
&  $M_{V}({\rm mag})$       & $-$23.00  & $-$21.98  & $-$20.98  & $-$20.04  & $-$18.99  & $-$17.97  & $-$16.96 \\
&  $M_{G}(10^{9}M_{\odot})$ &   812  &   291  &   108  &   42.0  &   14.8  &   5.35  &   1.98 \\
&  $T_{G}$(Gyr)  &  15.00  &  15.00  &  15.00  &  15.00  &  15.00  &  15.00  &  15.00 \\
{\it metallicity} &  $t_{gw}$(Gyr)  &  0.515  &  0.320  &  0.235  &  0.185  &  0.147  &  0.120  &  0.100 \\
{\it sequence} & $<\log Z/Z_{\odot}>$ & $-$0.014  & $-$0.116  & $-$0.213  & $-$0.304  & $-$0.404  & $-$0.502  & $-$0.596 \\
&  $\log <Z/Z_{\odot}>$ & 0.147 & 0.025 & $-$0.079 & $-$0.173 & $-$0.273 & $-$0.368 & $-$0.458 \\
&  $M / L_{B}$     &  8.406  &  7.568  &  6.846  &  6.240  &  5.652  &  5.144  &  4.700 \\
&  $U - V$      &  1.671  &  1.596  &  1.521  &  1.452  &  1.378  &  1.308  &  1.242 \\
&  $V - K$      &  3.355  &  3.273  &  3.192  &  3.116  &  3.032  &  2.949  &  2.868 \\ \hline\hline
&  $M_{V}({\rm mag})$       & $-$23.00  & $-$22.00  & $-$21.00  & $-$20.00  & $-$19.00  & $-$18.00  & $-$17.00 \\
&  $M_{G}(10^{9}M_{\odot})$     &   812  &   232  &   68.5  &   22.2  &   6.20  &   1.86  &  0.693 \\
&  $T_{G}$(Gyr)  &  15.00  &  9.97  &  6.88  &   5.57  &   3.68  &   2.54  &   2.30 \\
{\it age} &  $t_{gw}$(Gyr)  &  0.515  &  0.515  &  0.515  &  0.515  &  0.515  &  0.515  &  0.515 \\
{\it sequence} & $<\log Z/Z_{\odot}>$ & $-$0.014  & 0.002  &  0.013  & 0.004  &  0.013  &  0.060  &  0.039 \\
&  $\log <Z/Z_{\odot}>$ & 0.147 & 0.160 & 0.176 & 0.160 & 0.166 & 0.207 & 0.189 \\
&  $M / L_{B}$     &  8.406  &  5.856  &  4.190  &  3.334  &  2.266  &  1.635  &  1.483 \\
&  $U - V$      &  1.671  &  1.571  &  1.465  &  1.408  &  1.326  &  1.211  &  1.146 \\
&  $V - K$      &  3.355  &  3.274  &  3.194  &  3.113  &  3.033  &  2.952  &  2.871 \\ \hline\hline
   \end{tabular}
   \end{center}
 \end{table*}

 \begin{table*}
   \begin{center}
   \caption{Model comparison}
   \label{table-3}
   \begin{tabular}{c|ccccccccccc}\hline
 &  $M_{V}$  & $x$ & $t_{gw}$ & $\tau$ & $\tau_{in}$ &
$T_{G}$ & $M_{G}$ & $<Z>$ & $V-K$ & $U-V$ & $M/L_{B}$ \\ \hline\hline
this & $-$22.49 & 1.20 & 0.39 & 0.10 & 0.10 & 15 & 485 & 0.025 &
3.314 & 1.633 & 7.974 \\
work & $-$16.96 & 1.20 & 0.10 & 0.10 & 0.10 & 15 & 1.98 & 0.007 &
2.868 & 1.242 & 4.700 \\ \hline
Bressan et al. & $-$22.56 & 1.35 & 0.31 & 0.14 & 0.10 & 15 & 648 &
0.031 & 3.280 & 1.588 & 9.682 \\
(1996) & $-$16.82 & 1.35 & 0.43 & 1.00 & 0.10 & 15 & 2 & 0.008 &
2.793 & 1.215 & 5.265 \\ \hline
   \end{tabular}
   \end{center}
 \end{table*}

\section{Comparison with observations}

\subsection{Age-metallicity degeneracy}

 Figures 2 and 3 compare E1 and E4 models 
 to E1 and E4 spectral templates (see Arimoto 1996 for the details).
 E1 model has parameters $T_{G}=15$ Gyrs and $t_{gw}=0.52$ Gyrs.
 In Fig.2a, the lower solid line gives E1 model and the upper one
 shows the E1 template, respectively.
 Residuals between the two spectra are shown in Fig.2b.
 E1 model reproduces the E1 template very well except for
 $\lambda < 3000\AA$.

 Two E4 models (E4M and E4A) are confronted with the E4 template.
 E4M is one of the metallicity sequence with
 $T_{G}=15$ Gyrs and $t_{gw}=0.17$ Gyrs
 (lower solid line in Fig.3a).
 E4A is taken from the age sequence with $T_{G}=4.0$ Gyrs and
 $t_{gw}=0.52$ Gyrs (dotted line in Fig.3a).
 Both E4M and E4A models have almost identical spectrum
 and give excellent fits to the E4 template except for
 $\lambda=3000-4000\AA$, where the
 observational data might have systematic errors (the E4 template
 spectrum in this wavelength range is indeed an amalgam of several
 different galaxies taken from different sources; see Arimoto 1996).
 This possible error is suggested from the fact that the E4 template is
 0.2 magnitude bluer in $U-V$ compared to Bower et al.'s (1992b)
 regression line, as seen in Fig.1b.

 From Figs. 2 and 3,
 one cannot tell which is the key factor, age or metallicity,
 that brings a systematic difference in the spectral energy distributions
 between the E1 and E4 templates. This is the so-called
 {\it age-metallicity degeneracy}.

 To disentangle the degeneracy,
 luminosity and colour evolution of distant
 cluster ellipticals are crucial.
 Figs. 4a and 4b show the luminosity and colour evolutions
 of E1, E4M, and E4A models in the observer's frame. Filters are chosen for
 Gunn's system $g$ ($\lambda_{\rm eff}=4930\AA$) and $r$
 ($\lambda_{\rm eff}=6550\AA$) bands.
 Transmission functions are taken from Thuan \& Gunn (1976).
 Photometric zeropoints are defined by Thuan \& Gunn (1976) in such a way
 that the star BD$+17^{\circ}4708$ of spectral type F6V has $g=9.5$ and
 $g-r=0.0$,
 and its spectral energy distribution is obtained by K.Shimasaku
 (private communication).
 Adopted cosmological parameters are $H_{0}=50$ km s$^{-1}$
 Mpc$^{-1}$ and $q_{0}=0.1$.

 E4A model brightens rapidly in $r$-band as a function of redshift
 from $z=0$ to $z \simeq 0.24$, beyond of which E4A model dims abruptly
 as it is approaching to the formation epoch $z_{f} \simeq 0.26$ (Fig.4).
 $g-r$ of E4A model is kept nearly constant to $z \simeq 0.16$
 but becomes significantly bluer at higher redshifts since a galaxy is
 mainly composed of young massive stellar populations. Evolutionary behaviour
 of E4M model is almost similar to E1 model which becomes fainter
 and redder almost monotonically towards higher redshifts (Fig.4) as a 
 result of spectral reddening by redshift which is larger than
 intrinsic bluing of spectral shape by galaxy evolution.
 Such a conspicuous difference between E4A and E4M models clearly indicates
 that the colour and luminosity evolution must have definitive informations
 for {\it breaking} the age-metallicity degeneracy.

 We note that $g-r$ of E1 and E4M models
 in the observer's frame become redder
 in $0 \leq z \leq 0.35$ due to the passage of $g$-band
 through $4000\AA$ break, remain nearly
 unchanged in $0.35 \leq z \leq 0.7$,
 and become redder again in $z \geq 0.7$.
 Stagnation at $z \sim 0.5$ happens because a decrease of
 $g$-band flux is balanced by rapid dimming of $r$-band flux
 caused by its passage through CH G-band ($\sim 4300\AA$) to
 $4000\AA$ break; thus the feature is characteristic in Gunn's 
 $g-r$.

 \begin{figure}
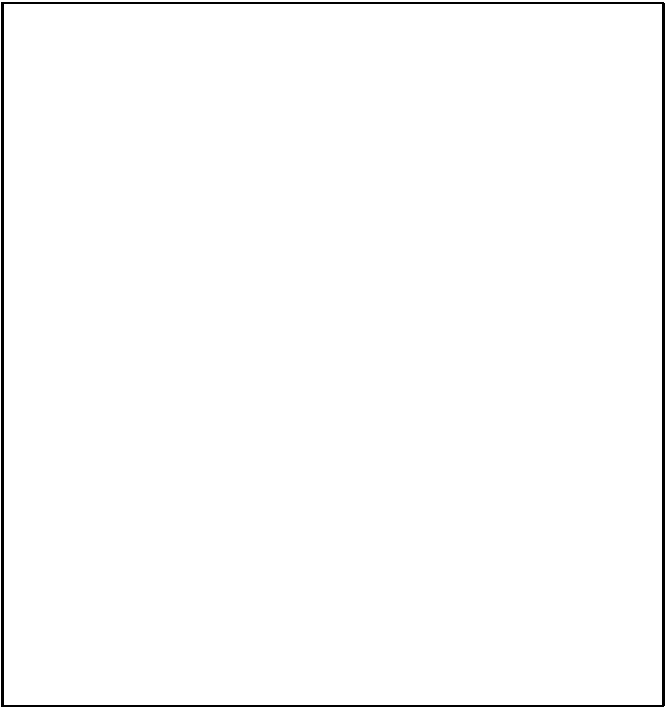

 \picplace{9.345cm}
 \caption{(a) Comparison of E1 model with the E1 spectrum template.
  Filled circles correspond to the fluxes at the effective wavelengths of
  $J$, $H$ and $K$ bands.
  (b) Residuals of the fluxes between E1 model and E1 template
  (obs $-$ model).}
 \label{fig-2}
 \end{figure}

 \begin{figure}
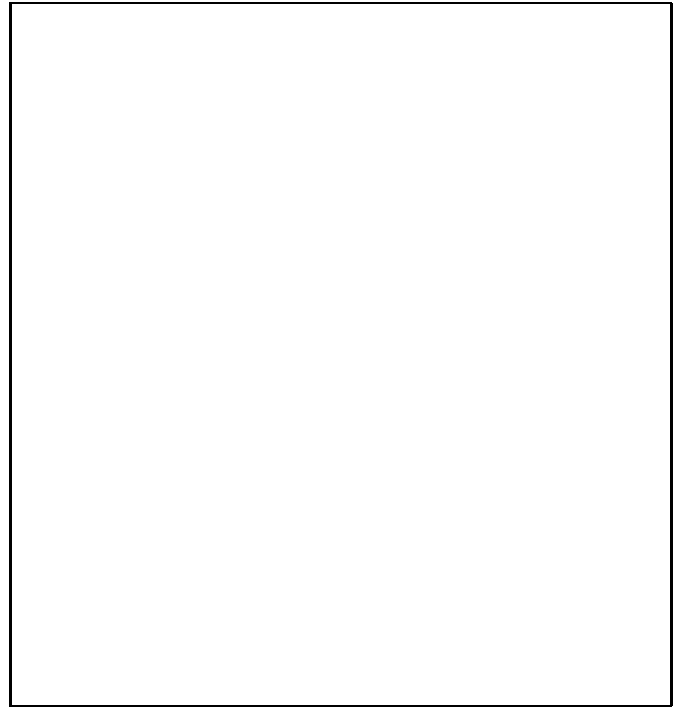

 \picplace{9.345cm}
 \caption{(a) Comparison of E4M (lower solid line) and E4A (lower dotted line)
  models with the E4 spectrum template.
  (b) Residuals of the fluxes between E4M (solid line and filled circles)
  and E4A (dotted line and open circles) models and E4 template
  (obs $-$ model).}
 \label{fig-3}
 \end{figure}

 \begin{figure}
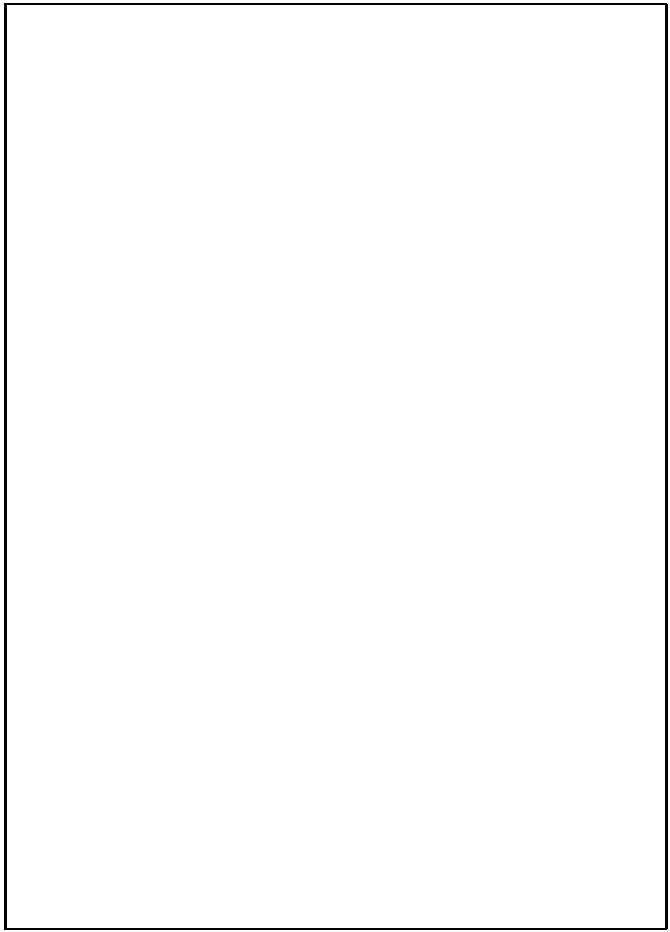

 \picplace{12.3cm}
 \caption{(a) $r-$luminosity evolution of E1, E4M, and E4A models in the
 observer's frame. $H_{0}=50$ km s$^{-1}$ Mpc$^{-1}$ and $q_{0}=0.1$.
 (b) The same as Fig.4a, but for $g-r$ colour evolution.}
 \label{fig-4}
 \end{figure}

 \subsection{Distant clusters}

 Evolutions of the CM relation in $g-r$ vs. $M_{r}$ diagram are
 presented in Figs. 5 and 6 for the metallicity sequence and the age sequence,
 respectively. Cosmological parameters are again
 $H_{0}=50$ km s$^{-1}$ Mpc$^{-1}$ and $q_{0}=0.1$.
 The age of the universe is then about 16.5 Gyrs, and
 a galactic age $T_{G}=15$ Gyrs
 corresponds to the formation redshift $z_{f} \simeq 5.4$, and $T_{G}=2.3$
 Gyrs to $z_{f} \simeq 0.14$.
 Crosses on each line indicate the position of each model defined at
 $z=0$. Galaxy masses at $z=0$ are indicated for some models.

 Observational data of E/S0 galaxies in two distant clusters
 Abell 2390 ($z=0.228$; triangles) and Abell 851 ($z=0.407$; circles)
 are superposed in Fig.5.
 For Abell 2390, data are taken from Yee et al. (1996). The membership
 is confirmed by redshift measurement and E/S0 morphologies are
 classified spectroscopically.
 Galaxies in the central field of the cluster
 ($7.3\arcmin \times 9.1\arcmin$)
 are represented by filled triangles.
 Since photometric data by Yee et al. (1996)
 are taken by CCD detector which reduces blue light,
 the transmission functions should be
 different from those of original system of Thuan \& Gunn (1976).
 According to Schneider, Gunn \& Hoessel (1983),
 we convert their data to the original
 Gunn's system by using conversion formula given by
 Schneider et al. (1983) :
 $r=r_{c}+0.044(g-r)_{c}$ and $g-r=1.199(g-r)_{c}$, where subscript $c$
 indicates the CCD magnitude.
 For Abell 851, galaxies classified as E/S0
 with HST images are taken from Dressler et al. (1994).
 Filled circles indicate cluster members confirmed spectroscopically.
 Absolute magnitudes $M_{r}$ are calculated
 from the redshift of each cluster with a help of
 the adopted cosmological parameters.

 Both clusters Abell 2390 and Abell 852 show well-defined CM relations of E/S0
 galaxies over 3 $\sim$ 4 magnitude range, and that these CM relations are
 almost in parallel to the CM relation at $z=0$ which is calculated from model
 sequences calibrated with the CM relation of Coma in $V-K$ vs. $M_{V}$
 diagram.  
 This trend is very well reproduced by the metallicity sequence (Fig.5).
 In the observer's frame, the CM relation evolves almost in parallel towards
 redder colours and is well defined even at $z=1$. 
 Indeed, Stanford, Eisenhardt and Dickinson (1995) 
 recently analyzed the
 optical$-K$ CM relation of elliptical galaxies in Abell 370 ($z=0.374$) 
 and Abell 851 ($z=0.407$) and showed that these two clusters have
 almost identical slopes and
 dispersions of the CM relations to those of Coma ellipticals.
 They therefore concluded that the CM relations of the two clusters
 are fully consistent with a picture of old passive evolution with age 
 more than 10 Gyrs.
 
 Due to the reason given at
 the end of \S 3.1, the CM relations at $z=0.4$ and $z=0.6$ are nearly
 identical. This is the characteristic in this photometric system and there
 will be no wonder if clusters in this redshift interval do not show any sign
 of evolution in their CM relations.
 As pointed out by Yee et al. (1996) and Dressler et al. (1994),
 their photometric data have uncertainties of the zeropoint
 calibration, which amount to $\sim$ 0.1 mag, but the slopes of CM relation
 should be much reliable.

 For the age sequence, on the contrary, the CM relation changes drastically
 (Fig.6). Galaxies {\it finally} on the CM relation at $z=0$ become rapidly
 brighter and bluer as a function of lookback time, because
 smaller galaxies become considerably young as they approach to their
 formation epoch. For example, galaxies smaller 
 than $2.2\times10^{10}M_{\odot}$ at $z=0$ no longer exist in the
 universe at $z=0.4$. As a result, at $z=0.2$,
 the CM relation holds only for 2.4 mag from the brightest end,
 and virtually disappears at $z \ge 0.6$.
 We note that the slope of the CM relation becomes progressively steeper
 towards higher redshifts, which is in high contrast to the metallicity
 sequence whose CM relation keeps its slope nearly constant from $z=0$ to 1.

 The CM relations of the two clusters Abell 2390 and Abell 851 extend to
 much fainter magnitudes than the theoretical loci predicted by the
 age sequence models. This is a clear evidence
 that the CM relation cannot be simply attributed to the 
 age difference alone.
 Thus, it can be concluded that the metallicity is the key factor that
 accounts for the CM relation of elliptical galaxies.

 Our conclusion does not depend on the cosmological parameters chosen.
 Figure 7 shows the evolution of CM relation of age sequence computed
 with $H_{0}=50$ km s$^{-1}$ Mpc$^{-1}$ and $q_{0}=0.5$.
 Here the age of the brightest galaxies are assumed to be 12 Gyrs instead
 of 15 Gyrs because the universe is younger ($\simeq 13$ Gyrs).
 $T_{G}=12$ Gyrs corresponds to $z_{f} \simeq 4.5$, which is nearly
 the same as that of the brightest galaxy in Fig.6. Galactic wind
 epoch $t_{gw}$ is set to 1.25 Gyrs.
 The situation is worse than the 
 $H_{0}=50$ km s$^{-1}$ Mpc$^{-1}$ and $q_{0}=0.1$ cosmology;
 the CM relation is deformed even at lower redshift,
 The same situation is realized when smaller formation redshift, 
 such as $z_f=1$ or 2, is adopted for a given set of cosmological parameters.

 \begin{figure*}
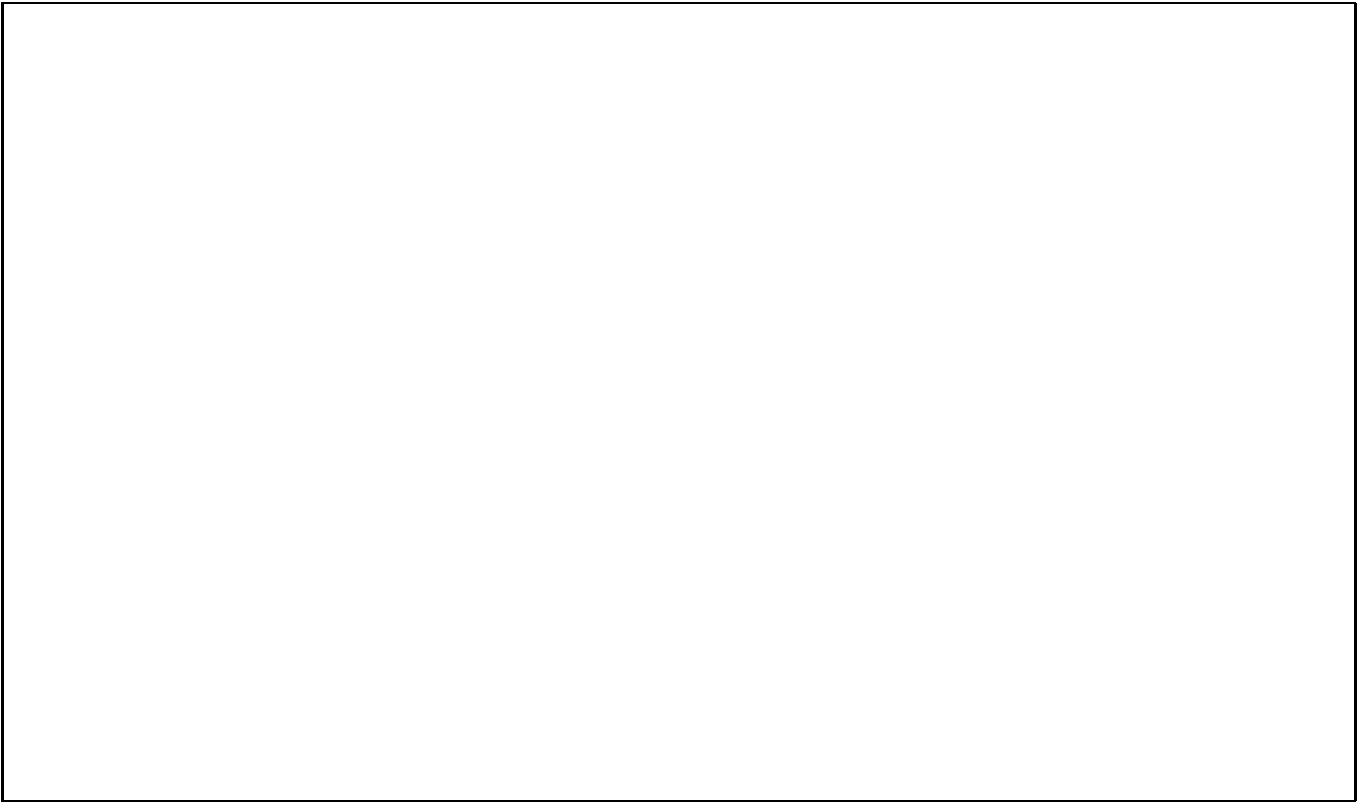

 \picplace{10.6cm}
 \caption {Evolution of the CM relation for the metallicity sequence in
 the observer's frame (solid and dashed lines). Triangles and circles
 indicate E/S0 galaxies in Abell 2390 ($z=0.228$) and Abell 851
 ($z=0.407$) clusters, respectively. Two filled squares give the E1
 (left) and the E4 (right) templates, respectively.}
 \label{fig-5}
 \end{figure*}

 \begin{figure*}
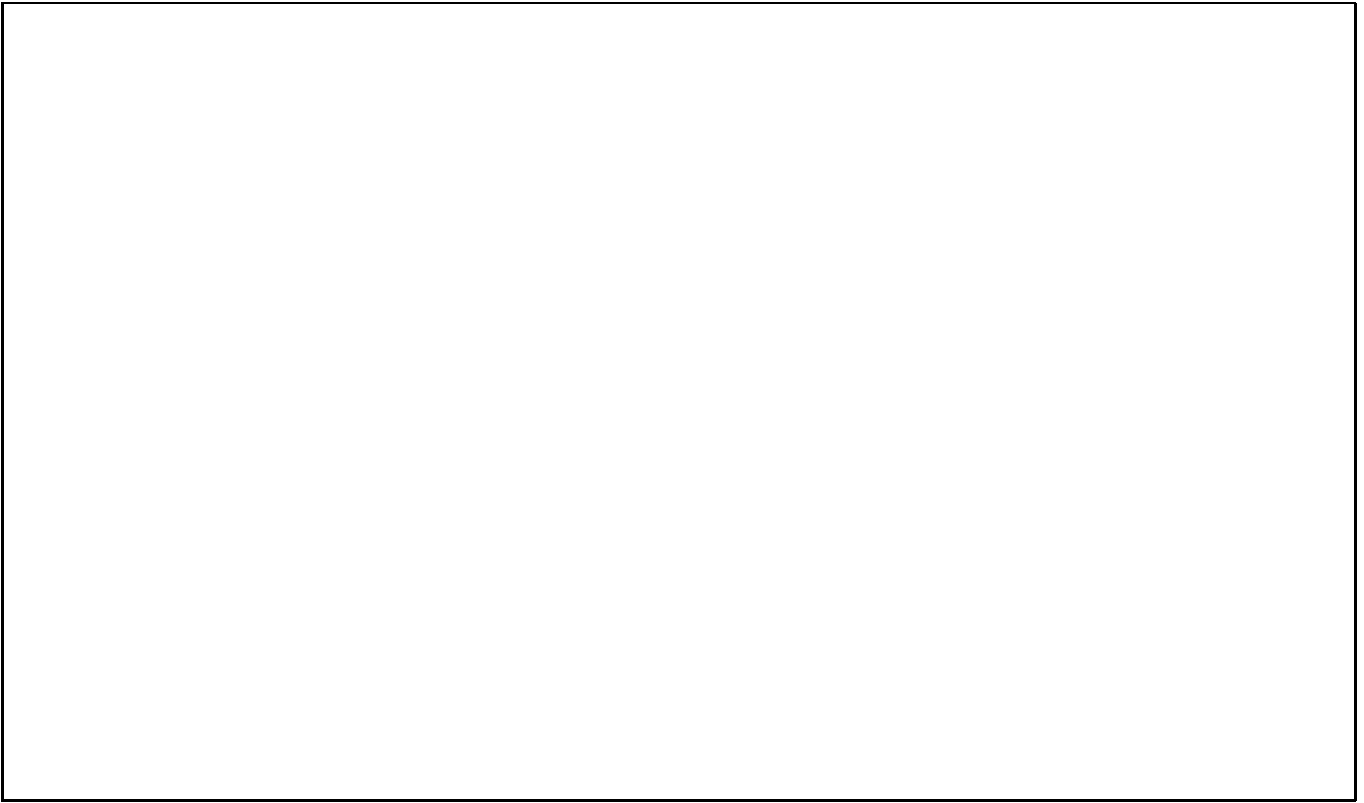

 \picplace{10.6cm}
 \caption{The same as Fig.5, but for the age sequence.}
 \label{fig-6}
 \end{figure*}

 \begin{figure*}
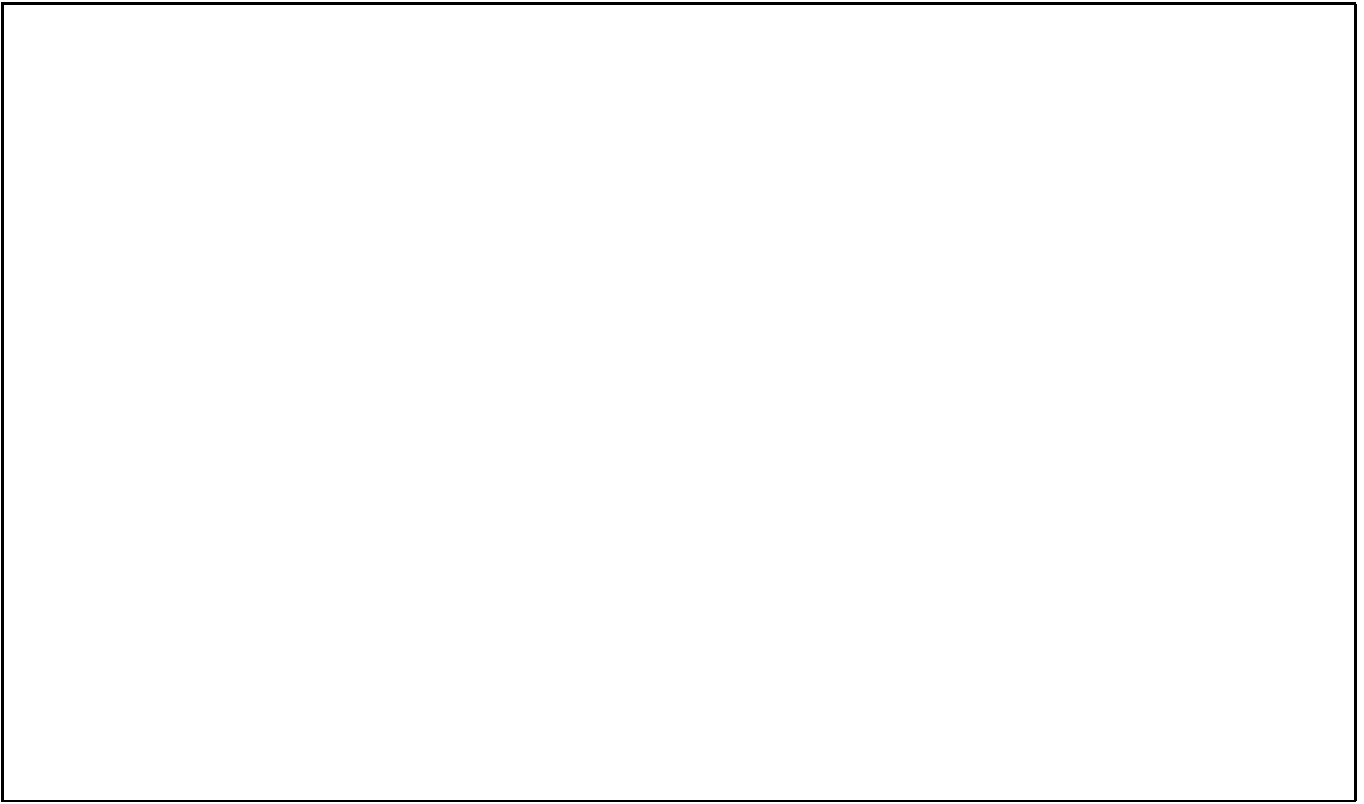

 \picplace{10.6cm}
 \caption{The same as Fig.6,
  but with the alternative cosmology ($H_{0}=50$ km s$^{-1}$ Mpc$^{-1}$,
  $q_{0}=0.5$).}
 \label{fig-7}
 \end{figure*}

\section {Discussion}

  Galaxies defining
 the CM relation at high redshifts may not be simply the progenitors
 of those defining the CM relation at $z \simeq 0$, since galaxies undergoing
 star formation at moderate to high redshifts could well enter the CM relation
 at lower redshifts, or alternatively, since galaxies in the CM relation at
 high redshift may leave it if they undergo new episode of star formation
 (e.g., Barger et al. 1996). However, if galaxies suffered star formation
 at moderate and high redshifts and if the light of the intermediate-age
 stellar populations dominate the luminosities of the present day ellipticals,
 these galaxies should deviate considerably from the CM relation, unless
 the star formation episodes occurred beyond a redshift of 2 
 (Bower et al. 1992b). In other words, the light from the intermediate-age
 stars cannot be a primary cause of the CM relation of cluster ellipticals
 today. Figure 5 shows that the CM relations were already established
 in Abell 2390 at $z=0.228$ (in particular, E/S0 in the central field)
 and Abell 851 at $z=0.407$. This implies that these galaxies had evolved
 {\it passively} down to these redshifts, and if the CM relations at these
 redshifts are due to the age effect, the epoch of galaxy formation must
 have been very precisely tuned for each specified galaxy mass. We believe
 this is very unlikely. 

 Based on a hierarchical clustering model of galaxy formation,
 Kauffmann (1995) demonstrated that some elliptical galaxies could be assembled
 and enter the CM relation even from $z=0.4$ to now, but she also
 indicated that most of the stars even in such elliptical galaxies
 formed fairly in early epoch such as $z > 1.9$ and the resulting CM
 relation could keep the small dispersion of Bower et al. (1992b)
 (Kauffmann 1996). Kauffmann (1995) did not obtain the CM relation for
 ellipticals, probably because she did not include any metal enrichment
 in the models. She mentioned that it is likely that the observed CM 
 relation is due to higher degree of metal enrichment in luminous
 galaxies.
 
 As suggested by dynamical disturbances in elliptical galaxies such as
 shells/ripples, multiple nuclei, and rapidly rotating and/or
 counter rotating cores (e.g., Kormendy \& Djorgovski 1989),
 or by the intermediate-age stellar populations suggested by
 the optimizing population synthesis (O'Connell 1976; 
 Pickles 1985b; Rose 1985; O'Connell 1986; Bica 1988; Rose et al. 1994),
 some ellipticals, especially those in less dense environment, must have
 undergone recent star formation, possibly by interaction with other galaxies.
 Schweizer \& Seitzer (1992) analized E/S0 galaxies in the field and
 groups and found that their $U-B$ and $B-V$ become systematically bluer
 as the amount of fine structure increases, which
 might spread the colour dispersion, though still small, compared to
 those of cluster ellipticals. This in turn suggests that the interaction
 with other galaxies causing these dynamical disturbances cannot be the 
 major cause of the CM relations in clusters, as otherwise the CM relations
 should exhibit much larger dispersions than Bower et al. (1992b) found. 

 In distant clusters ($z=0.31$) also, Barger et al. (1996) showed that
 upto 30 per cent of cluster members could undergo recent star bursts
 based on the high fraction of H$\delta$-strong galaxies,
 though many of them are regular spheroidals ({\it i.e.,} E/S0s).
 But we think these recent star formation should be secondary,
 with bulk of stars formed in early epoch,
 otherwise it must violate the small dispersion of Bower et al (1992b).
 Thus, some ellipticals might turn blue away from CM relation owing to the
 recent secondary star formation and come back again to the relation
 later, but in terms of the origin of CM relation, we could well push
 back the formation epoch far into the past.

 The results obtained in this paper are in excellent agreement with other,
 independent lines of evidence of an early completion of star formation in
 elliptical galaxies. The tightness of the correlation of galaxy colours 
 with central velocity dispersion sets indeed an upper limit of $\sim 2$
 Gyrs for the age dispersion of the bulk of stars in ellipticals 
 (Bower et al. 1992b). An identical upper limit to the age dispersion of 
 such galaxies is also established by the small dispersion on the 
 fundamental plane (Renzini \& Ciotti 1993). The most plausible interpretation
 of these evidences is that the bulk of star formation in cluster 
 ellipticals was basically completed at $z \ge 2$, which is also suggested
 by the Mg$_2$ vs velocity dispersion relation of elliptical galaxies 
 in Abell 370 ($z=0.37$) (Bender, Ziegler, \& Bruzual 1996).
 Fully consistent with
 this picture is also the detection of {\it passive} evolution of
 cluster ellipticals out to $z \simeq 1$, that seemingly requires a formation
 redshift in excess of $\sim 2$ (Arag\'on-Salamanca et al. 1993; Dickinson
 1996). Finally, we note that the Fe II (UV+optical)/Mg II $\lambda 2798$
 flux ratios of quasars B1422+231 at $z=3.62$ (Kawara et al. 1996) and 
 PKS 1937-101 (Taniguchi et al. 1996) at $z=3.79$ indicate that the 
 majority of stars in the host galaxies of these quasars, presumably 
 ellipticals, have formed much earlier than these redshifts.

 Fairly good agreements between the theoretical CM relations and the
 empirical ones of the two clusters are encouraging us to conduct a
 detailed study of cluster CM relations at intermediate and high redshifts.
 A confrontation of the model and the HST photometry of about dozen
 clusters will be given in a subsequent paper (T.Kodama, N.Arimoto, 
 A.Arag\'on-Salamanca, \& R.S.Ellis, 1996, in preparation).

 Finally we should comment on our definition of the epoch of galaxy formation.
 Quite often the epoch of galaxy formation is defined to be the time when
 the protogalaxy decouples from the Hubble expansion to the formation of
 the dark halo of the galaxy. However, this process is virtually invisible
 and difficult to observe. Instead, as is often done by observational
 cosmologists, we identify the epoch of galaxy 
 formation with observable events, or more precisely, the time when
 protogalaxies (normal massive ellipticals and bulges of spiral galaxies)
 are undergoing their first major episode of star formation. 
 Then, the arguments given in this paper suggest that stars formed during
 this episode, at $z \ge 2$, dominate the light of the elliptical galaxy
 and the CM relation was established at that epoch. Since only the galactic
 wind model has so far been successful in explaining the observed CM relation,
 we are led to reach a conclusion that the CM relation takes its origin
 at early times from galactic wind feedback.
 However, we should also mention that an alternative scenario of galaxy
 formation, the hierarchical clusrtering model discussed by Kauffmann (1995),
 could well satisfy the observational constraints we have used here.
 Indeed, Tinsley \& Larson (1979) were able to reproduce the metallicity--mass
 relation of the form $Z_s \propto M_s^{0.25}$ (Tinsley 1978) with a model
 of hierarchical sequence of mergers among subsystems in which stars are
 assumed to form in bursts induced by galaxy--galaxy collisions. However,
 possible age effect on line indices which Tinsley (1978) used to derive
 the stellar metallicity was entirely ignored. The age effect on the CM
 relation was fully taken into account in Kauffmann's (1995) model, but the
 stellar metallicity effect was ignored instead. Therefore, there is no
 successful hierarchical clustering model that can reasonably reproduce
 the observed CM relation. Wether or not the hierarchical clustering model
 can account for the CM relation is yet to be fully investigated.

\section {Conclusions}

 Evolutionary models for elliptical galaxies are constructed
 by using a new population synthesis code.
 The dissipative collapse picture by Larson (1974) is 
 adopted and model parameters are
 adjusted to reproduce the CM relation of Coma ellipticals.
 Two evolutionary sequences are calculated under the contexts of {\it
 metallicity hypothesis} and {\it age hypothesis}, both can equally
 explain the CM relation at the present epoch.
 The confrontation with the observational CM diagrams of E/S0 
 galaxies in the two distant clusters Abell 2390 ($z=0.228$) and
 Abell 851 ($z=0.407$) rejects the {\it age hypothesis}, and strongly suggests
 that the bulk of stars were formed early in elliptical galaxies with
 a possibly minor
 contamination of the intermediate-age stars,
 and give a reinsuring confirmation of many previous contentions
 that the CM relation takes its origin at early times probably 
 from galactic wind feedback. 
 This conclusion does not depend on the IMF nor the SFR,
 cosmological parameters, neither.

\begin{acknowledgements}

 We are grateful to H.Saio for providing us a stellar evolution code,
 and to S.Yi and P.Demarque for the evolutionary tracks of horizontal branch
 stars. We also thank to our anonymous referee for her/his constructive
 suggestions. T.K. thanks to the 
 Japan Society for Promotion of Science (JSPS)
 Research Fellowships for Young Scientists.
 N.A. is grateful to PPARK (UK) for financial support for his stay at the 
 Institute of Astronomy, University of Cambridge, and thanks to 
 A.Arag\'on-Salamanca and R.S.Ellis for intensive discussions.
 This work was financially supported in part by a Grant-in-Aid for the
 Scientific Research (No.06640349) by the Japanese Ministry of Education,
 Culture, Sports and Science.

\end{acknowledgements}

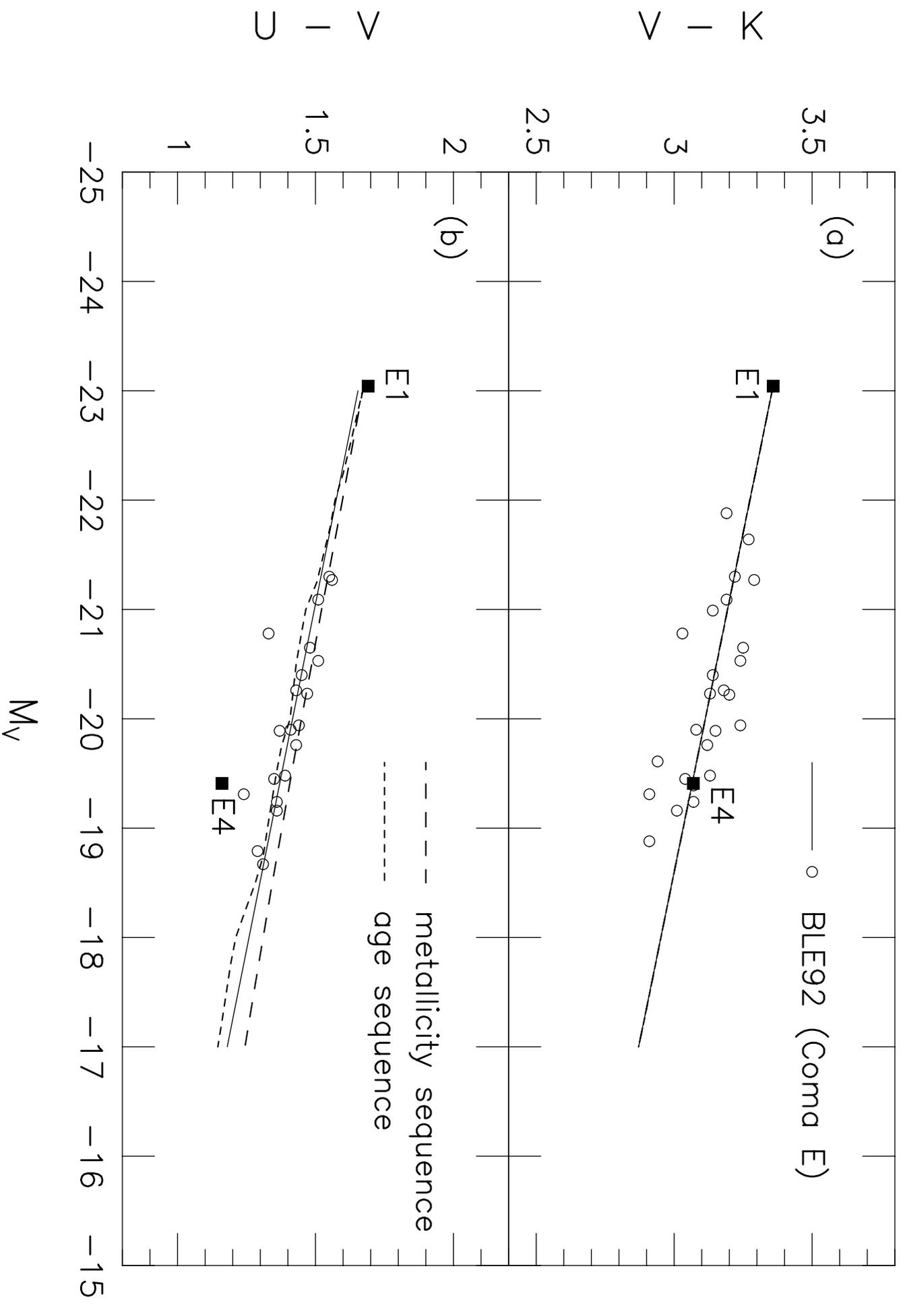
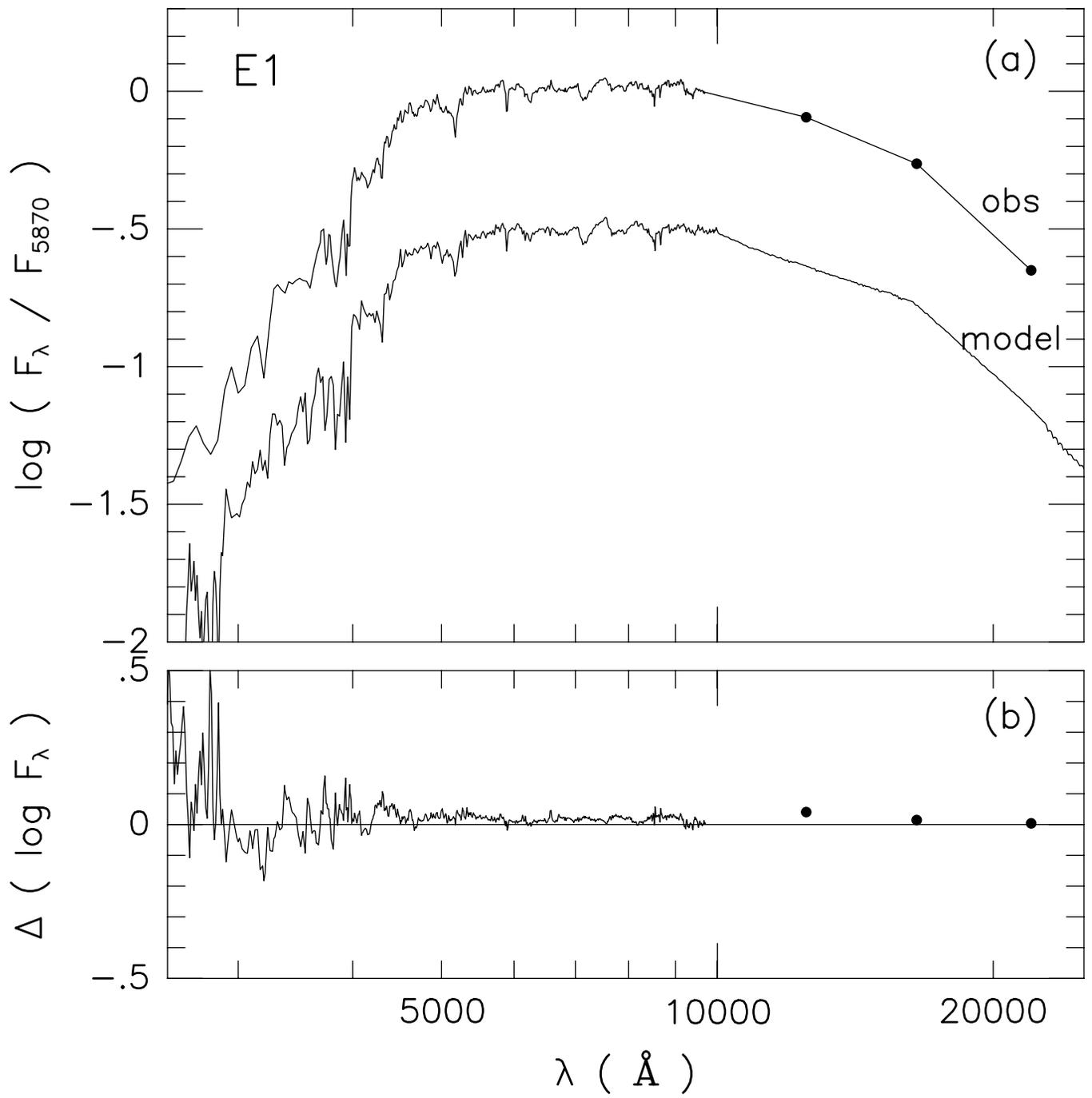
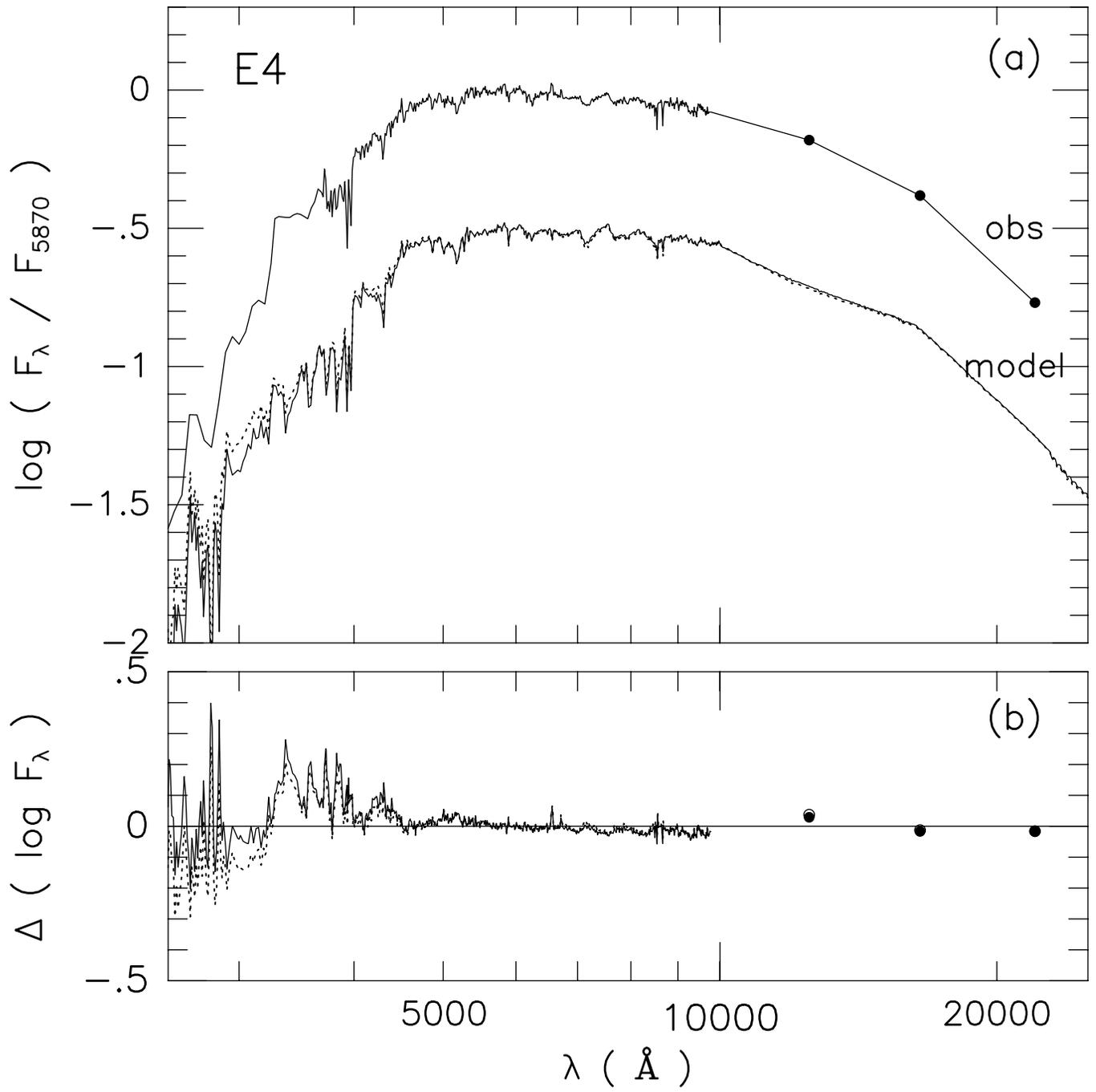
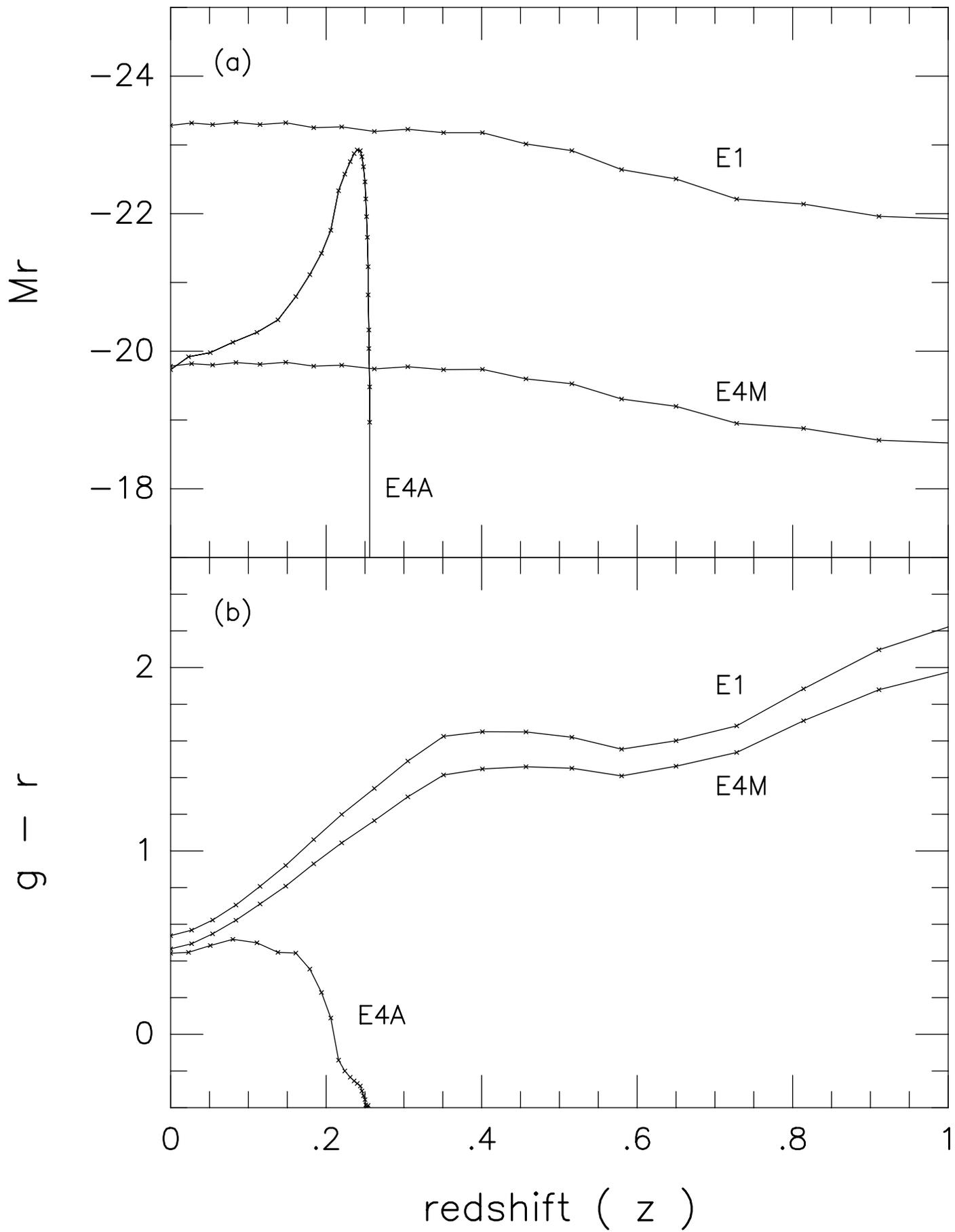
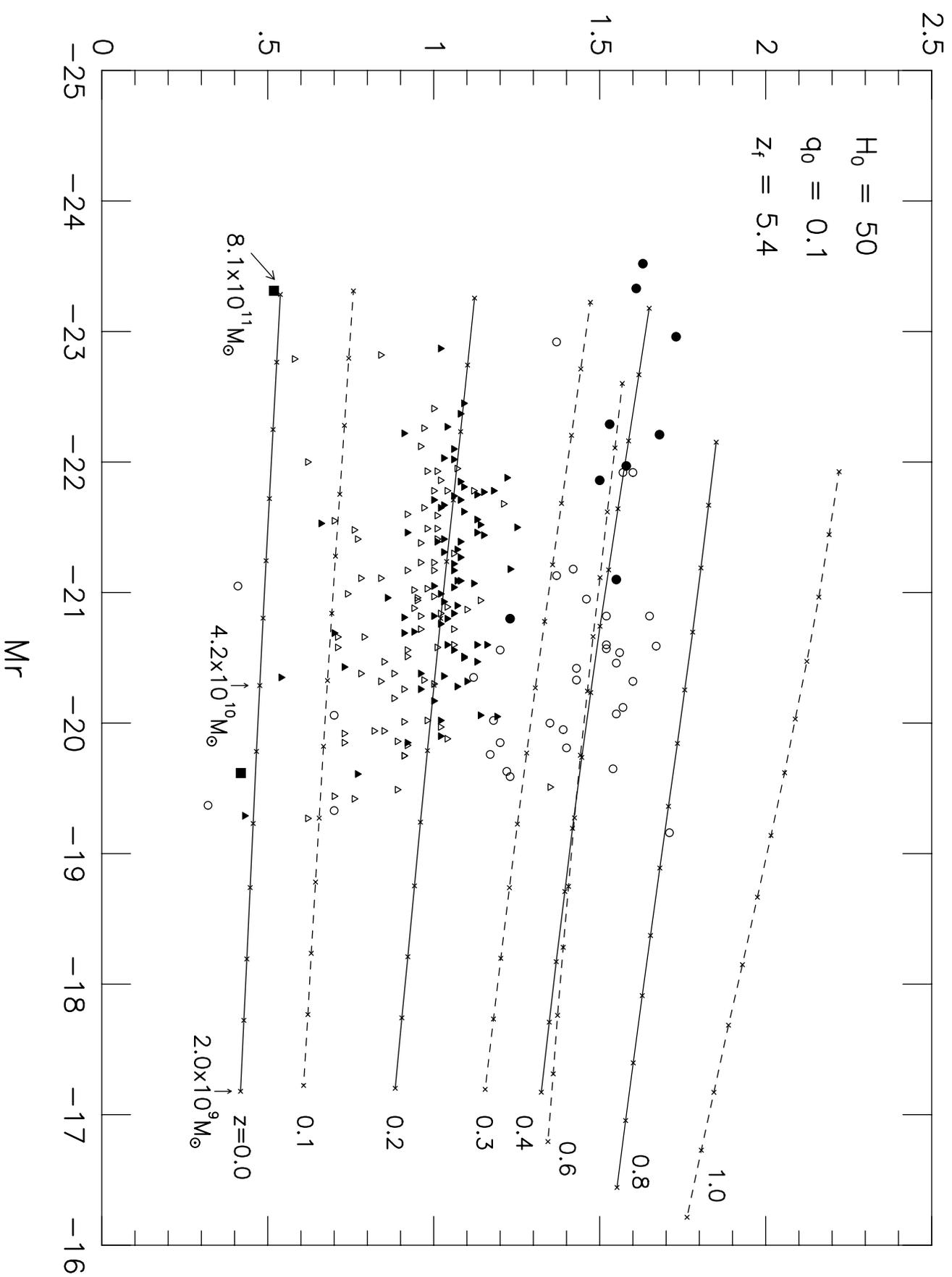
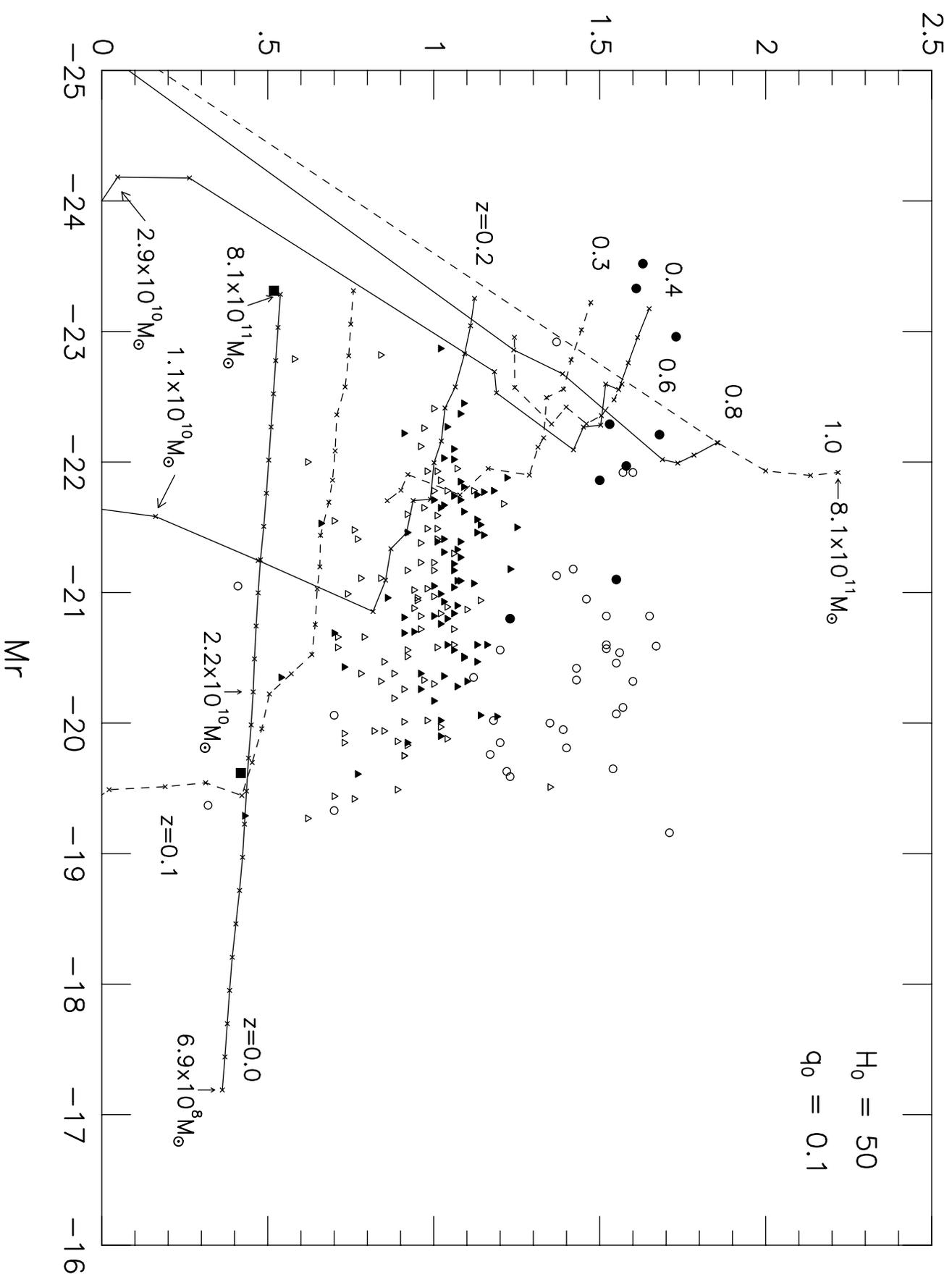
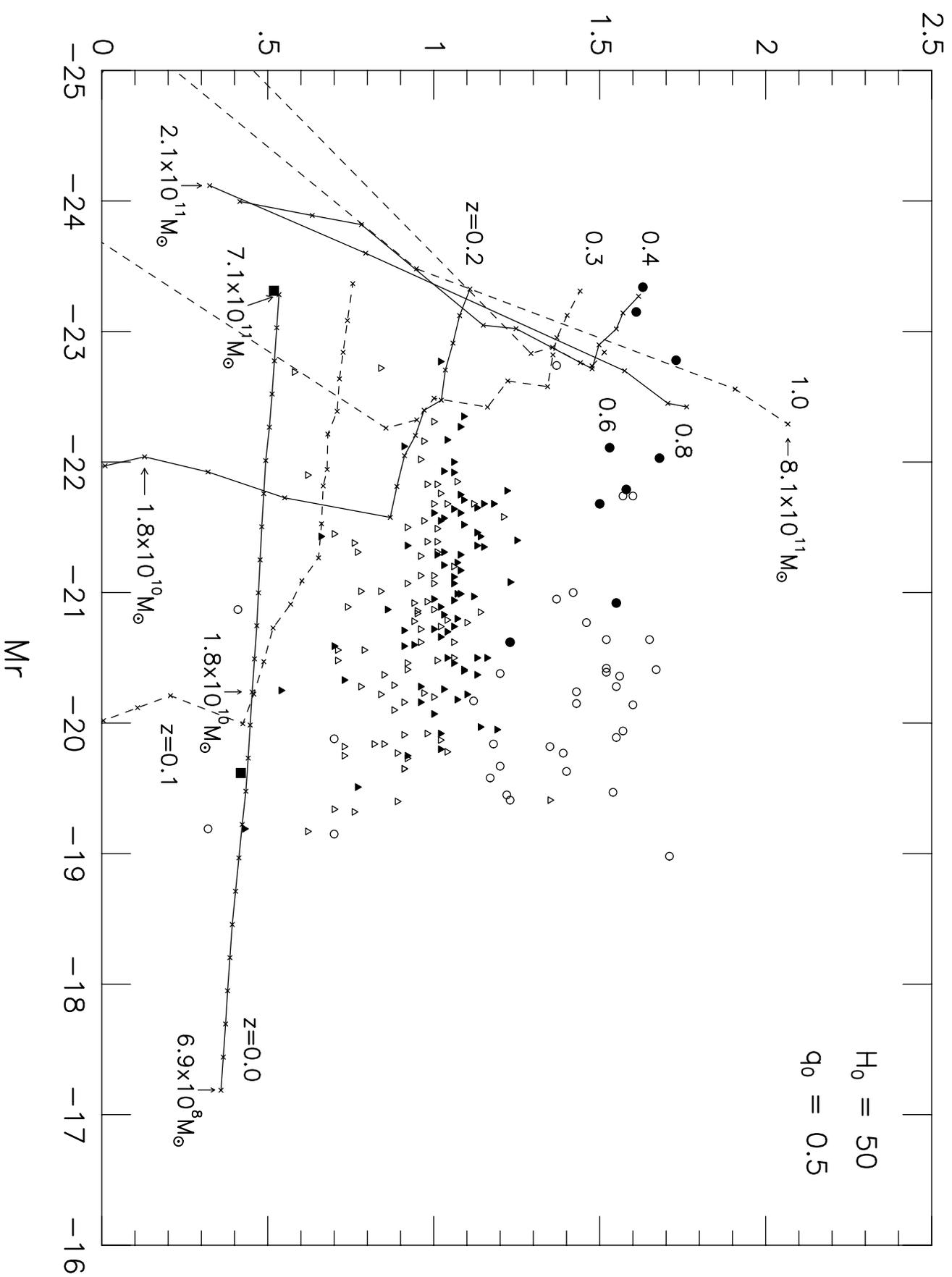
\end{document}